\documentclass[12pt]{iopart} 
\usepackage{graphicx} 
\usepackage{amssymb} 
\usepackage{color}
\begin{document}
\title[Reducing False Alarms for Compact Binary Coalescences Searches]{Methods for Reducing False Alarms in Searches for Compact Binary Coalescences in LIGO Data}
\author{J. Slutsky, L. Blackburn, D. A. Brown, L. Cadonati, J. Cain, M. Cavagli\`a, S. Chatterji, N. Christensen, M. Coughlin, S. Desai, G. Gonz\'alez, T. Isogai, E. Katsavounidis, B. Rankins, T. Reed, K. Riles, P. Shawhan, J. R. Smith, N. Zotov, J. Zweizig}

\begin{abstract}
The LIGO detectors are sensitive to a variety of noise transients of non-astrophysical origin.  Instrumental glitches and environmental disturbances increase the false alarm rate in the searches for gravitational waves.  Using times already identified when the interferometers produced data of questionable quality, or when the channels that monitor the interferometer indicated non-stationarity, we have developed techniques to safely and effectively veto false triggers from the compact binary coalescences (CBCs) search pipeline. 
\end{abstract}

\pacs{04.30.Tv, 04.80.Nn, 07.60.Ly }
\maketitle

\section{Introduction}

In October 2007, the Laser Interferometer Gravitational-wave Observatory (LIGO)~\cite{Abbott:2007kv} detectors completed a fifth data run over a two-year long period, denoted as S5, during which one year of triple-coincidence data was collected at design sensitivity.  During S5, LIGO consisted of two interferometers in Hanford, Washington and one in Livingston, Louisiana. In Hanford, the interferometers had arm lengths of 4 km and 2 km, referred to as H1 and H2, respectively.  In Livingston, there was a single 4 km interferometer, referred to as L1.  The LIGO detectors were sensitive to the coalescences of massive compact binaries up to 30 Mpc for neutron stars~\cite{Collaboration:2009tt,Abbott:2009qj}, and even farther for binary black holes.  Ongoing ``Enhanced'' LIGO upgrades, as well as the future ``Advanced'' LIGO upgrades, are planned to increase the sensitivity significantly.  

The LIGO Scientific Collaboration (LSC) performs astrophysical searches for gravitational waves from compact binary coalescences (CBCs), including the inspiral, merger, and ringdown of compact binary systems of neutron stars and black holes.  These searches use matched filtering with template banks that include a variety of durations and frequency ranges for inspirals or ringdowns.  When a LIGO interferometer is locked and operational, data are recorded in what it is called ``science mode.'' In order to confidently make statements on astrophysical upper limits and detections from science mode data, characterization of the LIGO detectors and their data quality is vital. The LIGO detectors are sensitive to a variety of noise transients of non-astrophysical origin, including disturbances within the instrument and environmental noise sources. Triggers generated by these disturbances may occur at different times and with different amplitudes for each template, increasing the false alarm rate in the searches for gravitational waves.

In this paper we discuss techniques for vetoing non-astrophysical transient noises effectively, and thereby reducing their effect on searches for gravitational waves.  These methods were developed on searches for low mass CBCs in the first of the two years of S5 as described in~\cite{Collaboration:2009tt,Abbott:2009qj}, though they are applicable to future searches as well.

In Section 2 we present the broad range of data quality issues that were present in LIGO data. In Section 3 we briefly review the search methods for which these techniques were developed. The techniques that we have so far developed and implemented to evaluate vetoes are explained in Section 4. The categorization and application of these vetoes is described in Section 5. In Section 6 we describe proposed methods to extend and automate our vetoes for use in future CBC searches.  We present our conclusions in Section 7.

\section{Data Quality Studies}
There are two broad categories of spurious transients in LIGO data: instrumental and environmental noises.  Within these two classes of noise sources, there are dozens of identified phenomena that require vetoing.  LIGO records hundreds of channels of data on the state of internal degrees of freedom of the interferometers, as well as the output from environmental sensors located nearby.  

When a set of transients effect the stability or sensitivity of the gravitational wave data, members of the Detector Characterization and Glitch groups within the LIGO Scientific Collaboration (LSC) work to determine the source of these transients, using the auxiliary channel data~\cite{Blackburn:2008ah}.  When a given noise source is identified, it is documented with sets of time intervals referred to as data quality flags, which begin and end on integer GPS seconds. These data quality flags are discussed in Sections 2.1 and 2.2.

Alternatively, one can begin with the hundreds of data channels, and search for correlations between the outputs of some algorithm on these channels and the gravitational wave data channel.  The ``auxiliary channel'' vetoes used by the CBC search are described in Section 2.3.

\subsection{Data quality flags from Instrumental noise transients}  

Each LIGO interferometer makes an extremely sensitive comparison of the lengths of its arms. This necessitates the ability to sense and control minute changes in displacement and alignment of the suspended optics, as well as intensity and phase in the laser~\cite{Abbott:2007kv}.  The control systems have both digital and analog components, as well as a variety of complex filtering schemes. Instrumental noise transients, or {\it glitches}, sometimes correspond to fluctuations of large amplitude and short duration in the control systems.  While these may be prompted at times by environmental effects, they can be identified and vetoed by using the control channels alone, as they are well known failure modes of the control systems.  We describe some examples in the following paragraphs.  

\begin{itemize}

\item{Overflows.} The feedback control signals used to control the interferometer arm lengths and mirror alignments are processed and recorded in digital channels. When the amplitude of such a signal exceeds the maximum amplitude the channel can accommodate, it ``overflows", and the signal abruptly flattens to read as this maximum value, until the quantity falls back below this
threshold.  This discontinuity in the control signal usually introduces transients at the time of the overflow.  

\item{Calibration line dropouts.} Signals of single frequency are continuously injected into the feedback control system to provide calibration. Occasionally,  these signals ``drop out" for short periods of time, usually one second.  This discontinuous jump in the control signal produces artifacts in the data both when the calibration line drops out and when it resumes.

\item{Light scattering.} The two interferometers at the Hanford site share the same vacuum enclosure.  During times when one of the interferometers was locked and in science mode, and the other was not locked, the swinging of the mirrors of the unlocked interferometer scattered light into the locked interferometer.  This produced strong, short duration transients, though not necessarily overflows.  

\item{Arm cavity light dips.} Brief mirror misalignments caused drops in the power in the arm cavities, and thus transients in the data.  

\end{itemize}

\subsection{Data quality flags from environmental noise transients}  
Environmental noise transients correspond to the coupling of mechanical vibrations and
electromagnetic glitches that enter into the interferometer. Seismic motion, human activity near
the LIGO sites, and weather are the most common sources of mechanical vibrations.  Similarly to instrumental transients, we describe some examples that have been identified by the LIGO detector characterization group in the following paragraphs. 

\begin{itemize}

\item{Electromagnetic disturbances.} The electronic systems of the interferometers are susceptible
to electromagnetic interference, due both to glitches in the power lines and electronics noise at both sites.  Magnetometers arrayed around the detector are used to diagnose these signals.

\item{Weather-related transients.} The Hanford site is arid, with little to block the
wind from pushing against the buildings housing the interferometers.  Times at Hanford with wind
speeds above 30 MPH are problematic. Weather and ocean waves also contribute to ground motion in the frequency range 0.1 Hz to 0.35 Hz at both sites, particularly Livingston.

\item{Seismic disturbances.} Seismic activity from different noise sources have different characteristic frequencies. Earthquakes around the globe introduce transient noise in the frequency range 0.03 Hz to 0.1 Hz.  Nearby human activities such as trucks, logging, and trains, produce disturbances with frequencies greater than 1 Hz, and can be so extreme that the interferometers often cannot stay locked.  Even when they remain locked, significant noise transients frequently occur.
\end{itemize}
The ground motion described in the latter two bullets above occurs at frequencies below a few Hz, while the frequency range the interferometers are most sensitive in is from 100 Hz to 1000 Hz.  There is non-linear coupling from the ground motion into the interferometers, which results in increased glitching at higher frequencies during times of high ground motion, and for this reason, these environmental effects are especially important.

\subsection{KleineWelle triggers from auxiliary channels}
Data quality flags identify times affected by specific issues and use auxiliary channels as appropriate to locate the problems.  Rather than starting with a known or plausible noise coupling, we search over hundreds of the auxiliary channels for transients that are coincident between any of these channels and the gravitational wave channel.  This information can be used both for producing veto intervals, and for finding clues about the problems left unflagged.  

A wavelet based algorithm, KleineWelle (KW)~\cite{Chatterji:2004qg}, is used to analyze interferometer control and environmental data. It is a valuable source of triggers for detector characterization because of its low computational cost allowing it to be applied to many data channels.  The algorithm produces trigger lists contains the peak time and significance of the trigger.  During S5, KW analyzed the gravitational wave channel and a variety of important auxiliary channels for the three LIGO detectors, including interferometer channels used in the feedback systems of the detectors and channels containing data from the environmental monitors.

\section{Searches for Gravitational Waves from CBCs}
Matched filtering is the optimal method of finding known signals in data with stationary
Gaussian noise~\cite{wainstein:1962}.  The searches for CBCs in S5 data used a matched
filter method to compare theoretically predicted waveforms with the LIGO gravitational wave channel data~\cite{Allen:2005fk}.  Because the masses of the components of the binary determine both the duration and the frequency profile of the gravitational radiation, template banks consisting of many different waveforms are used~\cite{Owen:1998dk}.  
\begin{figure}[ht!]
\begin{center}
\includegraphics[width=0.8\textwidth]{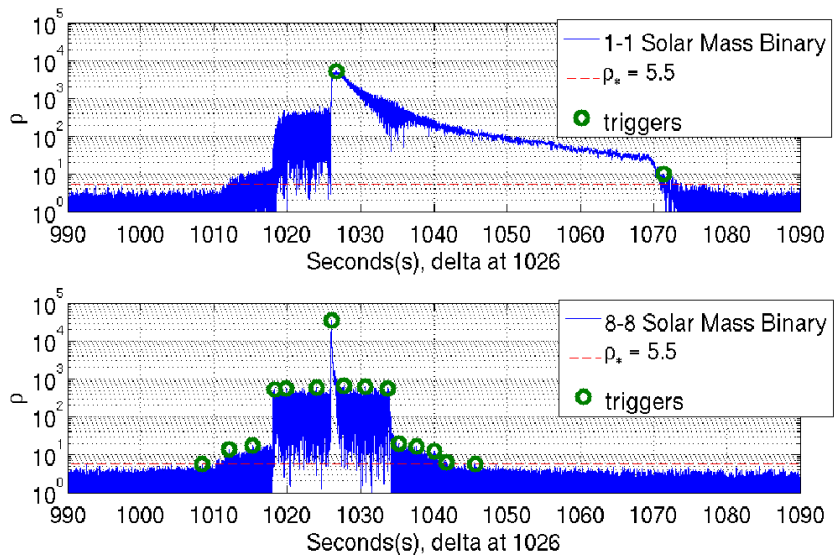}
\end{center}
\caption{Matched filtering an impulse in Gaussian noise, colored based on the LIGO design sensitivity, using the CBC search with a single template.  The result for two templates, top consisting of a binary of 1 solar mass objects, and bottom a binary of 8 solar mass objects.  The red dashed line is the search threshold $\rho_{*}$ of 5.5, and the green circles indicate the positions of the resulting triggers.
}
\label{fig:twoTemplates}
\end{figure}

When the signal to noise ratio $\rho$ of a template rises above a pre-set threshold $\rho_*$, a trigger corresponding to the time of coalescence for the binary system is recorded~\cite{Allen:2005fk}.  Single interferometer triggers are compared for coincidence in time and component mass parameters between the different interferometers, to produce coincident triggers. These coincident triggers are further subjected to signal consistency checks, for example a $\chi^2$ test in time/frequency bins between the template and the data~\cite{Allen:2004}.  In order to estimate the accidental rate of the coincident triggers, the searches perform ``time-slide'' analysis, in which the data from the interferometers are offset by time shifts large compared to the light travel time between the detectors, resulting in the production of coincident triggers that can only be of instrumental origin.  Knowing the accidental rate, we can estimate the significance of unshifted coincidences.  Those with the highest significance are examined as candidate events and, in the absence of a detection, used to calculate astrophysical upper limits~\cite{LIGOS3S4all,Collaboration:2009tt,Abbott:2009qj}. 

Transients of non-astrophysical origin, as described in Section 2, often produce triggers of large $\rho$, as there is significant power in these transients~\cite{LIGOS3S4Tuning}. Even with the signal consistency checks, disturbances of non-astrophysical origin increase the false alarm rate in the time-slides, as well as producing accidental coincidences from non-astrophysical events in the unshifted data.  This has the effect of reducing the significance of the loudest events which are not caused by these transients, as the rate of coincidences in the time-slides is increased, and thus the measured false alarm rate  of the events in unshifted time is elevated.  It therefore has the effect of ``burying'' good gravitational wave candidates, as coincidences due to transient detector noises can produce significant outliers.  In order to reduce these effects, we have learned to define time intervals within which triggers should not be trusted.  These are called {\it vetoes}.

The CBC searches use banks of templates, each starting from a low frequency cutoff of 40 Hz (defined by the detector sensitivity) and increasing in amplitude and frequency until the coalescence time of the represented system.  These templates have durations of up to 44 seconds for the lowest mass templates (binary of 1 solar mass objects).  

Figure \ref{fig:twoTemplates} depicts the filtering of an impulse in Gaussian noise, colored to match the LIGO noise spectrum, using the CBC search for a single template.  We show the result for two different templates, the top panel consisting of a binary of 1 solar mass objects, and the bottom panel a binary of 8 solar mass objects.  The dashed line is the search threshold $\rho_{*}$ of 5.5, and the circles indicate the positions of the resulting triggers.  The response of the matched filter search to loud impulsive transients in the data is complicated, with multiple simultaneous effects of the data and search code.  While the $\rho$ time series of both templates have a clear peak near the time of the impulse, the peaks do not perfectly overlap.  The lower mass template has a long tail down from the peak, and there is a plateau of high $\rho$ extending 8 seconds before and after the time of the impulse.  Additionally, while filtering with both templates results in a trigger with a large $\rho$ value, the higher mass template results in many more triggers.  While many of the triggers shown would fail the $\chi^2$ test mentioned above, any that even marginally pass this test will increase the rate of accidental coincidences in the time-slides, adversely effecting the significance calculation of unshifted coincidences. 
 
The difference in time and $\rho$ of the peak occurs because $\rho$ is recorded at the coalescence time of the matching template, and frequencies in each template are weighted by the frequency dependent noise spectrum of the detector.  The time of the top panel trigger is 0.66 seconds after that of the bottom panel, whereas their $\rho$ values are 5000 and 34000, respectively.  For a broadband transient, the time between when the transient occurs and when the coalescence time is recorded is determined by the time remaining in the waveform after its frequency content matches the sensitive frequency band of the detector (40 Hz to 1 kHz).  No transient is a perfect delta function, and there are many transient types that have different timescales and frequency content.  
The $\rho$ tail visible in the top panel of Figure \ref{fig:twoTemplates} occurs due to the aforementioned 44 second template duration.  When any part of the template is matched against the impulse, the $\rho$ is significantly above threshold.  Rather than record a trigger for all times with $\rho \geq \rho_{*}$, a trigger is only recorded if there is no larger value of $\rho$ within one template duration of its coalescence time.  The higher mass template therefore results in more triggers because its duration is of order 3 seconds, while the plateau is 16 seconds long.  The plateau is not caused by any attribute of the waveform, nor by the data adjacent the impulse, but rather is a phenomena intrinsic to the method used to estimate the power spectral density of the data.  This occurs in the presence of impulsive transients in the data, as described in sections 4.6 and 4.7 of Ref.~\cite{findchirp}.  Vetoes of impulsive transients for CBC searches must include this time.

In cases of extraordinarily powerful transients, we observe a second trigger slightly more than one template duration from the trigger at the peak of the $\rho$ timeseries, as is the case in the top panel of Figure \ref{fig:twoTemplates}.  Because the actual search is performed using a bank of thousands of templates of different durations, a significant number of triggers multiple seconds away from the impulse are recorded, as can be observed for the triggers with $\rho \sim100$ to the right of the peak $\rho$ triggers in Figure \ref{fig:H2trig}.  

For all these reasons, intervals containing transients of non-astrophysical origin often must be padded with extra duration to make them into effective vetoes for the CBC searches. This is done by examining the falloff of the triggers in $\rho$ before and after the transients in vetoed times, in order to include those triggers associated with the transient while working to minimize the deadtime by not padding more than necessary.  For the S5 searches, this was determined by examining plots of the maximum and median amplitude transients, as measured by the peak trigger $\rho$.  Efforts to automate this decision process are ongoing, as discussed in Section 5.2.  

Prior to S5, all CBC searches used a single set of veto definitions~\cite{Chatterji:2004qg,Vetoes}.  During S5,  CBC searches extended over a broad range of component masses.  The wide distribution in the template durations of the waveforms caused the triggers associated with transients in the data to appear at different times, and be sensitive to different frequency ranges.  Thus veto window paddings must be defined based on template waveforms included in each search, which are determined by the possible component masses of the binaries.

\section{Established Veto Techniques}
In this section we discuss techniques for using the aforementioned detector characterization work
to create vetoes for matched filter CBC searches.  We illustrate our methods with examples from CBC searches in the first year of LIGO's fifth science run, on which they were developed and implemented~\cite{Collaboration:2009tt,Abbott:2009qj}.  This work was done simultaneously and in consultation with the veto efforts for the searches for unmodeled gravitational wave bursts~\cite{Abbott:2009zi}.  The veto techniques discussed in this section are implemented by creating lists of times during which triggers from a search are suspected of originating not from gravitational radiation, but from instrumental or environmental disturbances.  

When deciding to use sets of time intervals as vetoes, we need to include all of the bad times of the interferometers, and as little of the surrounding science mode as possible.  Since not all vetoes are well understood, we need to create figures of merit, or metrics, to evaluate the effectiveness of the vetoes.  These veto intervals are  derived both from data quality flags (Section 4.3) and disturbances in auxiliary channels (Section 4.4).  We then classify these vetoes into categories (Section 5).

The {\it safety} of the veto intervals must also be ensured.  A veto is unsafe if it could be triggered by a true gravitational wave.  In order to insure that our instrumental vetoes are safe, we investigate their correlation with {\it hardware injections}, intentional transients introduced into the interferometer in order to properly tune the various searches in LIGO for gravitational wave signals.  The signals are injected directly into the gravitational wave channel itself, the differential arm length servo, to simulate the effect that a gravitational wave would have on the detector.  Since these hardware injections are intentional and controlled, they exist entirely within known time intervals.  

\subsection{Veto metrics}

Veto metrics were developed on single-interferometer triggers, from CBC searches in the first year of LIGO's S5 science run, as well as previous runs.  For this purpose, the triggers were clustered by keeping a trigger only when there are no triggers with larger value of $\rho$ within 10 seconds.  With this clustering, all the triggers from a single loud transient occurred in one, or at most two, clusters.  This had the effect of making the figures of merit independent of the number of waveforms in the template bank of the search.  Different searches may still use different clustering times, based on the different waveform durations.  A minimum $\rho$ of 8 for the clustered triggers was chosen in order to be sensitive to glitches that produce loud triggers from the template bank.  This threshold applies to the clusters used to measure the metrics, but the resulting vetoes are applied to all triggers falling in vetoed times.  

The percentage of triggers vetoed defines the {\it efficiency} of the veto $E = \frac{N_{vt}}{N_t}\cdot100\%$, where $N_{vt}$ is the number of clustered triggers vetoed and $N_t$ is the total number of clustered triggers.  If all outliers came from a single source, then the ideal veto would have $100\%$ efficiency, especially at large $\rho$.  In reality, there are many different sources of transient noise, as detailed in Section 2, and each may be responsible for only a few percent of the clustered triggers, and only in some specific range of $\rho$ values.  The efficiency quantifies the effectiveness of the veto for removing clusters, but more information is required to determine whether this removal is warranted.  Because the clustering is by loudest $\rho$, veto efficiency as a function of $\rho$ can be used to learn more about what population of clusters correspond to the transient event being vetoed.  

To determine the statistical significance of the efficiency, we need to compare it with the {\it deadtime}.
The percentage of science mode contained in a set of veto intervals defines the deadtime $D = \frac{T_v}{T}\cdot100\%$, where $T_v$ is the time vetoed and $T$ is the total science mode time, including the vetoed time.  If the veto only includes truly bad times, then by vetoing triggers within these times we are not reducing our chance of detection, as the noise transients already polluted this data.  In practice, the integer second duration of data quality flags, as well as the need for padded veto windows due to the nature of the CBC search (described in Section 4.2), limits how small the deadtime can be for veto intervals that remove common transient noises.  It also adds to the probability that a true gravitational wave event, occurring when the detectors are in science mode, will be missed if these times are vetoed.  Most vetoes have deadtimes of at least several tenths of a percent of the science mode time, although less understood or longer duration disturbances may lead to vetoes with larger deadtime percentages.  

Important to the determination of what vetoes are effective is the ratio of the efficiency over the deadtime.  Effective vetoes have a deadtime small compared to the efficiency, indicating that many more clusters are vetoed than one would expect by random chance.  This ratio is unity for ineffective vetoes, and large for effective vetoes, and can be expressed as
\begin{equation}
\label{eqn:EffDtRatio}
R_{ED} = \frac{E}{D} = \frac{N_{vt} \cdot T}{N_t \cdot T_v}.
\end{equation}

The percentage of veto intervals that contain at least one clustered trigger defines the {\it used percentage} such that $U = \frac{N_{wt}}{N_w}\cdot100\%$ where $N_{wt}$ is the number of veto windows that contain at least one cluster and $N_w$ is the total number of windows. For an ideal veto, every vetoed interval should contain at least one cluster, corresponding to one or more loud transient noises.  For vetoes with short time spans compared to the clustering time, it is more common to obtain values of the used percentage of less than 100\%, even for effective vetoes.  The statistical significance of this metric is made by comparing the used percentage for a veto to that expected if its intervals are uncorrelated with the triggers.  This expected used percentage is obtained by multiplying the length of a given veto interval $T_w$ by the average trigger rate, given by dividing the number of triggers by the available science mode time.  The ratio behaves similarly to $R_{ED}$, and can be expressed as 
\begin{equation}
\label{eqn:UsedPerRatio}
R_U = \frac{U}{T_w\frac{N_t}{T}\cdot 100\%} = \frac{N_{wt} \cdot T}{N_w \cdot T_w \cdot N_t} = \frac{N_{wt} \cdot T}{N_t \cdot T_v}.
\end{equation}
In the S5 run, there was a clustered trigger on average every 6 to 17 minutes, depending on interferometer. Effective vetoes have an expected used percentage small compared to the actual used percentage, indicating more intervals contain clusters than one would expect by random chance.

To evaluate the safety of each veto, the percentage of hardware injections that are vetoed is compared to the veto deadtime.  If veto intervals are correlated with the hardware injections, as indicated by a $R_{ED}$ for injections significantly greater than 1, the veto could be generated by an actual gravitational wave signal.  Such a veto is therefore ``unsafe'' and is not used.  In S5, zero data quality flags, and only one auxiliary channel, that were considered were found to be unsafe.

\subsection{Vetoes from data quality flags}
\begin{figure}[ht]
\begin{center}
\includegraphics[width=0.45\textwidth]{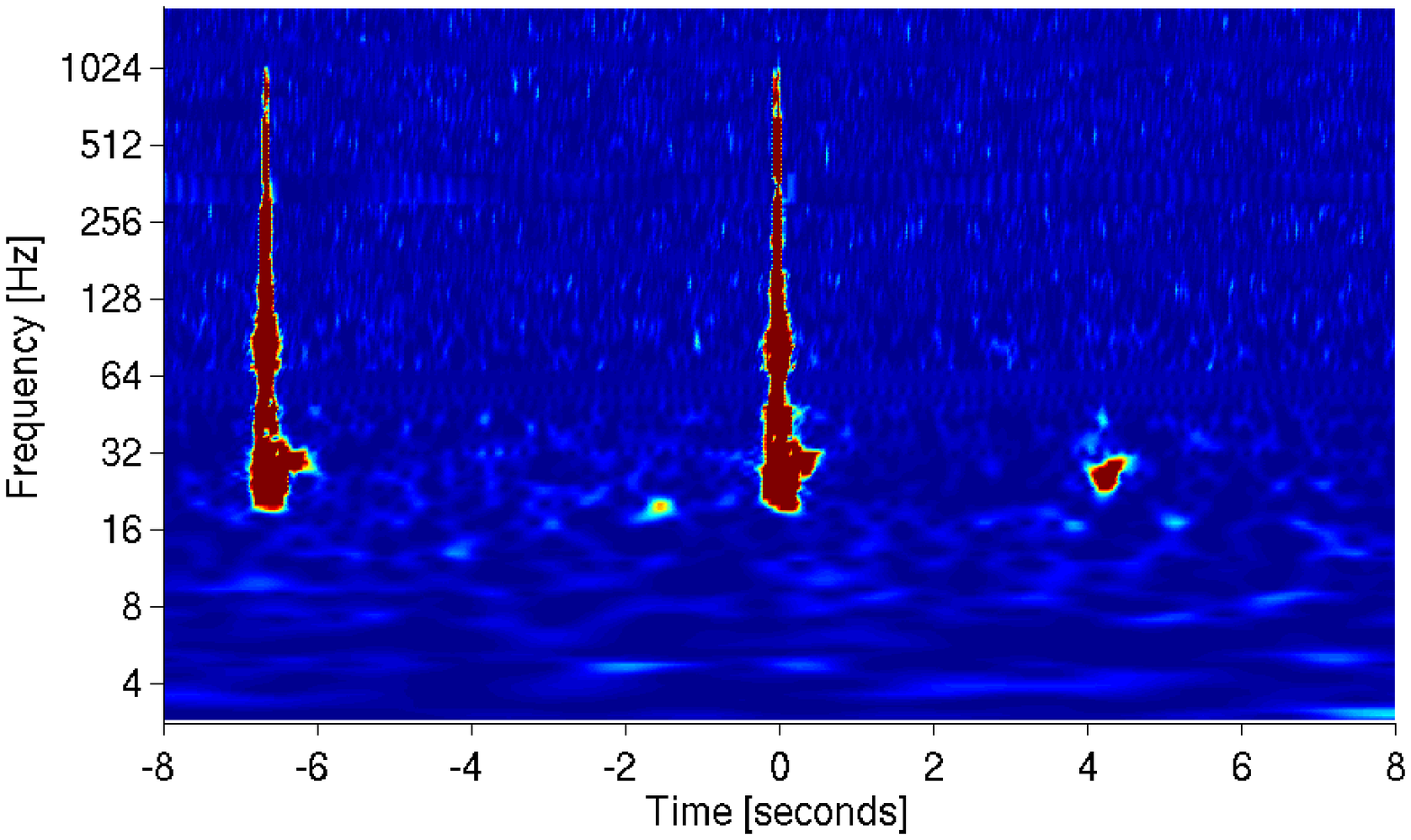}\includegraphics[width=0.45\textwidth]{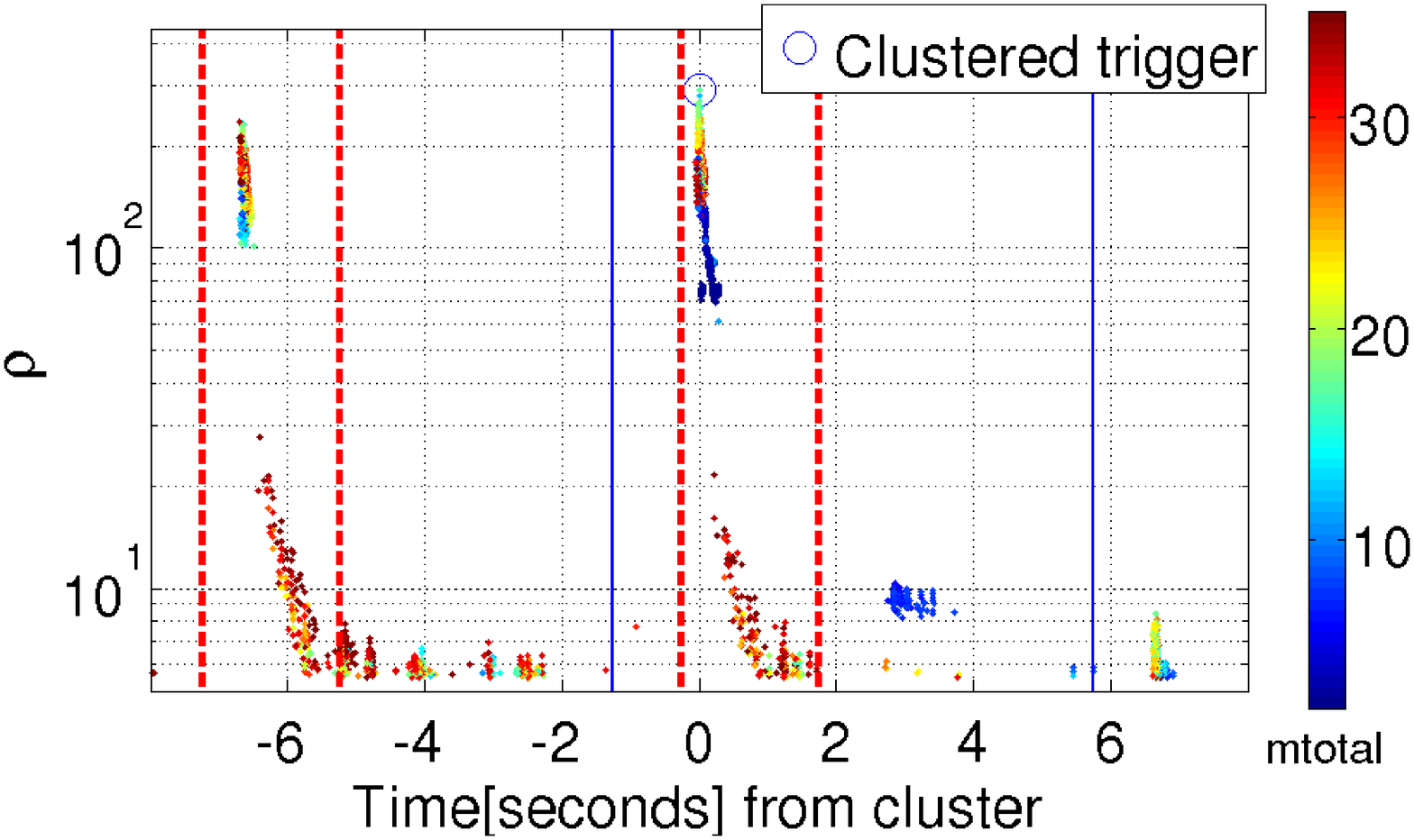}
\end{center}
\caption{A pair of overflows in the length sensing and control system of H2.  At left, a time-frequency representation~\cite{Chatterji:2004qg}.  At right, the effect of the transient on the production of unclustered triggers for the CBC search with total mass between 2 and 35 solar masses (dots), the 10 second clusters of these raw triggers (circle), the original Data Quality flags (dashed lines), and the expanded data quality veto after duration paddings are applied (solid lines).  }
\label{fig:H2trig}
\end{figure}

We apply our veto metrics and window paddings to data quality flags created by the Detector Characterization and Glitch groups within the LSC, as mentioned in Section 2.  A concrete example of a veto based on a data quality flag marking instrumental transients can be seen by examining the case of identified intervals containing an overflow in the length sensing and control loops for the H2 interferometer.  These overflows cause severe glitches, and are identified  within a second of their occurrence.  The overflows themselves are caused by other disturbances to the control systems such as seismic motion, but irrespective of the physical origin, the overflow itself produces  glitches in the gravitational wave channel.  Figure \ref{fig:H2trig} shows a time-frequency representation~\cite{Chatterji:2004qg} of the gravitational wave channel, as well as a plot of the unclustered triggers as a function of time, for the CBC search with total mass between 2 and 35 solar masses, around two typical transients caused by a type of overflow.  The data quality flag intervals start and stop on GPS seconds, and have a minimum duration of two seconds, centered around the times of the overflows, to ensure the glitches are not too close to the edges of the intervals. The loudest raw triggers occur near the transients, corresponding to the clustered trigger. Raw triggers subsequently fall off in $\rho$ over the next several seconds after the data quality interval.  

\begin{figure}[ht]
\begin{center}
\includegraphics[width=0.5\textwidth]{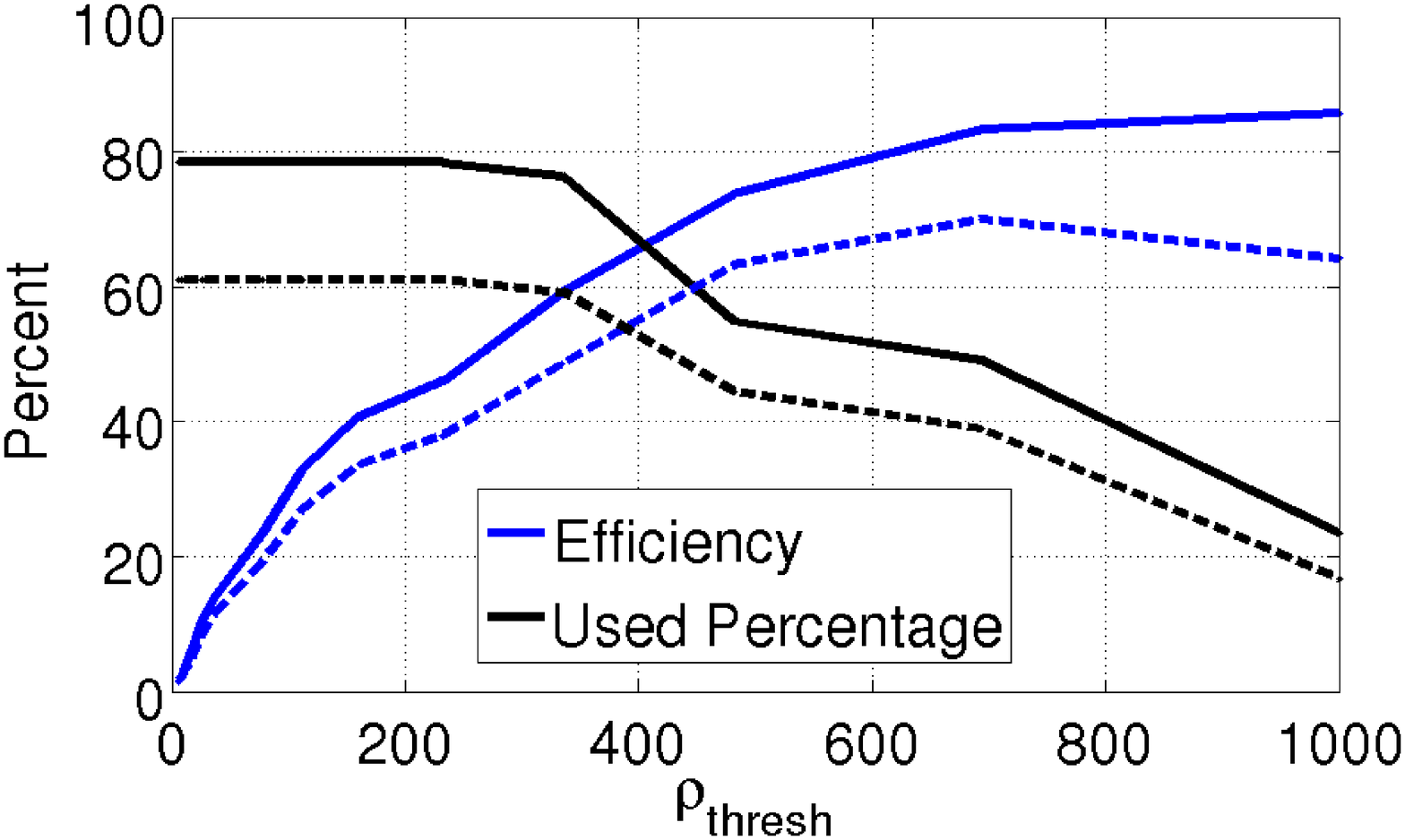}\includegraphics[width=0.5\textwidth]{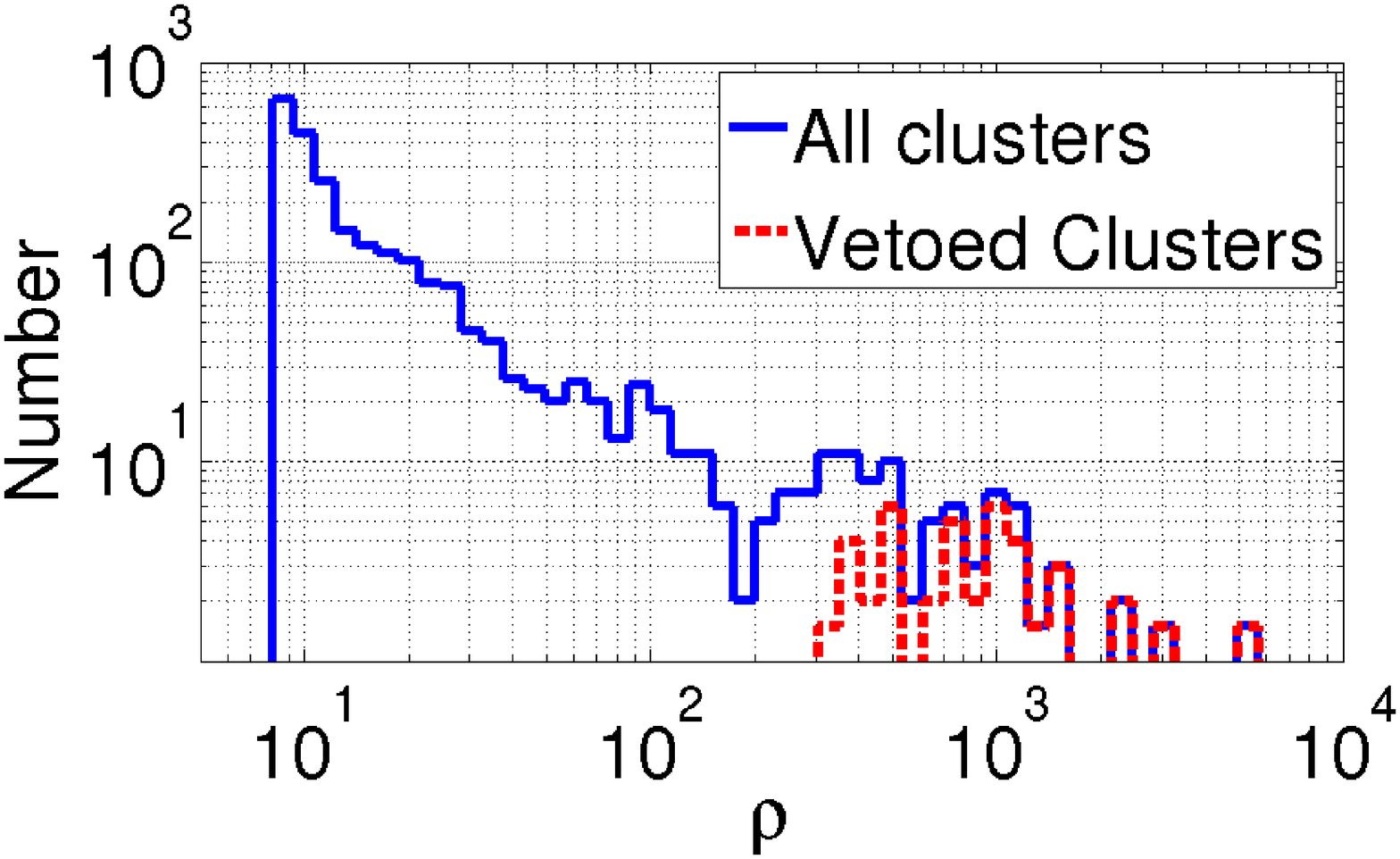}
\end{center}
\caption{The left plot shows the efficiency and used percentage as a function of a minimum
threshold single-interferometer cluster $\rho$ for the H2 Length Sensing and Control Overflow veto. The dashed lines are the values for the data quality flag before window paddings are added. The solid lines are the values after windows are added to veto the associated triggers.  The right plot
shows a log-log histogram for the same veto. All clusters found in science mode are shown in solid lines and vetoed clusters are shown in dashed lines.
}
\label{fig:H2eff}
\end{figure}

As shown in Figure \ref{fig:H2eff}, this flag has a used percentage of 62\% for a typical month during the first year of S5, indicating that these veto segments are well suited to vetoing triggers from transients.  These data quality veto segments have efficiency on all triggers of 1.4\%, and deadtime of 0.0037\%.  The ratio of the efficiency to the deadtime is more than 300.  This indicates a veto with a statistically significant correlation to the triggers, as the expectation for random chance would be a ratio of 1.  The efficiency is strongly dependent on the $\rho$ of the clustered triggers; for clusters with $\rho \geq 50$ it is 14\%, while for clusters with $\rho \geq 1000$, the efficiency is 64\%.  Two thirds of the loudest triggers found from the H2 interferometer in this month were due to overflows.

To attempt to account for as many triggers as possible associated with this transient, the veto interval is given 1 second of padding prior to the data quality interval, and 4 seconds after.  This alters the metrics, leading to a deadtime of 0.013\%, an efficiency for all clusters of 1.7 \%, a used percentage of 78 \%, and an efficiency to deadtime ratio of 130.  The expected used percentage from the trigger rate was only 0.58 \%, giving $R_U$ of slightly over 130.  For veto intervals with duration equal or less than the clustering time, only one cluster can be vetoed per interval, thus the number of veto windows used $N_{wt}$ approaches the number of triggers vetoed $N_{vt}$. Comparing Eqns \ref{eqn:EffDtRatio} and \ref{eqn:UsedPerRatio}, this means we expect $R_{ED} \approx R_U$, as we see in this example with a value of 130.  

For such loud transients, we expect all veto intervals to be used, but even for the loudest intervals we have tens of percent of the intervals that contain no cluster, as is the case with the aforementioned overflow flags.  Of the 22\% of the overflow veto intervals that are unused, 20 \% are within a clustering time of a clustered trigger.  These intervals, therefore, may well have many raw triggers, as the overflow on the lefthand side of Figure \ref{fig:H2trig} did, but only the overflow with the loudest raw trigger was marked by a clustered trigger.  Future efforts to automatically classify the effectiveness of vetoes (Section 5.1) would likely be sensitive to the anomalously low used percentages mentioned above, but for future searches the problem is significantly mitigated by employing a clustering window of 4 seconds rather than 10 seconds.  

The efficiency versus $\rho$ of the veto interval is shown in Figure \ref{fig:H2eff}.  This rises rapidly with the minimum $\rho$ of the clustered triggers, and the efficiency to deadtime ratio reaches 1300 for clusters with $\rho$ above 50 and over 5000 for clusters with $\rho$ above 500.   Efficiency and used percentage would be independent of $\rho$ if the times vetoed were random and uncorrelated with transient noises.

An example of a data quality flag for an environmental transient that we used to make a veto was the flag marking the times of elevated seismic motion due to trains passing through Livingston.  Early in S5, investigations of loud noise transients in L1 indicated that many such transients occurred in the minutes preceding loss of interferometer lock due to the passage of trains near the detector.  At each LIGO site, seismometers are located in each major building.  Since the trains pass closest to the end of the ``Y'' arm, the seismometer located there is most sensitive to the trains.  Specifically, the train-induced seismic motion was most pronounced along the direction of the arm in the 1-3 Hz frequency band.  Seismic disturbances due to trains were visible upon examining the 1-3 Hz band limited root mean square (BLRMS) value for minutes of the aforementioned seismometer channel.  Setting a minimum seismic threshold in that channel of $0.75\mu m/s$ in the 1-3 Hz BLRMS to identify times of passing trains, the two to three trains per day that passed the interferometer were identified.  Studies of the seismic motion induced by these trains compared with single interferometer online glitch monitoring codes~\cite{Blackburn:2008ah} showed correlation for up to a minute before and after the minute of peak seismic amplitude for each train.  Data quality flag vetoes were defined to mark these times. 
 
For this example, our metrics then yielded a deadtime of 0.69 \%, an efficiency of 2 \% for all clusters and 20 \% for clusters with $\rho_{thresh} \geq$ 100, and used percentage of 60 \% and 32 \% for each threshold respectively. $R_{ED}$ therefore increases from 3 to 30 with increasing $\rho_{thresh}$.  $R_U$ for the same values of $\rho_{thresh}$ increases from 0.78 to 16.  This veto is effective at eliminating a population of significant glitches, though not as loud or common as those from the overflows mentioned earlier.  Techniques for combining together the information from these data quality vetoes is discussed in Section 4.5.

\subsection{Auxiliary channel used percentage veto}
Even after taking all DQ flags into account, the number of triggers left unflagged is still very much in excess of those expected from random noise. Another approach is also possible, using the KW triggers that are generated on important auxiliary channels.    

The KW based veto method developed in S5 is similar to other KW based vetoes implemented in LIGO's S1, S2, S3 and S4  CBC searches~\cite{Vetoes,vetoGWDAW03}. Comparisons were made between KW triggers in interferometer and environmental channels and the clustered CBC triggers.  The KW significance can be used in the same way that $\rho$ is for CBC searches.  

For each channel, the threshold on KW significance was initially chosen to be above the background exponentially distributed triggers from Gaussian noise but low enough to catch noise transients. This threshold was then incremented until a used percentage for the veto of at least 50\% was achieved.  This was chosen to ensure that the times identified were likely to contain transients.  Veto intervals were then generated by taking intervals $\pm 1 s$ from the KW trigger times, and rounding away from the trigger to create intervals of $3s$ total duration.  Only channels that achieved a used percentage of 50\% were considered for veto use. 

In the tuning of the veto, already diagnosed problematic times, as described in Sections 5.1 through 5.3, were excluded from consideration.  Once each veto was defined, a list of time intervals to be excluded was created for all S5 Science mode data with the tuned parameters. The safety of each veto channel was determined using hardware injections, similar to the method used for data quality flags; those channels that produced a statistically significant overlap with hardware injection times were not implemented as vetoes. While the vetoes  were defined via comparison with the clustered inspiral triggers, the vetoes required padding to ensure that the unclustered triggers associated with each transient were also vetoed.  For channels that triggered coincident with large amplitude transients in the gravitational wave channel, we added $ 7 s$ veto window paddings to the beginning due to the plateau effect (Section 3) and end of the initial $3 s$ intervals.  For each CBC search these KW based used percentage vetoes were tuned and defined with respect to the single interferometer triggers from the specific analysis. 

In the first year of S5 there were a number of critical veto channels found in this manner. An important veto associated with an interferometer control channel was the feedback loop that keeps the H1 recycling cavity resonant. A veto based on environmental monitors came from the magnetometers located at the end of the Y-arm for L1.

\section{Veto categorization}

The goal of using vetoes is to reduce the false alarm rate, in order to more accurately assess the whether gravitational wave candidates are true detections.  Upon evaluating the available vetoes, we found that they do not all perform similarly, and divided them into categories.  Well understood vetoes have a low probability of accidentally vetoing gravitational waves, and significantly reduce the background.  More poorly understood vetoes can also reduce the background significantly, but with an increased chance of falsely dismissing actual gravitational waves.  We classified the vetoes into categories in order to allow searches to choose between using only the well established vetoes or aggressively using more poorly understood vetoes.  

Those vetoes classified as well understood almost always had higher $R_{ED}$ and $R_{U}$, as well as lower overall deadtime, than the less understood transients that correspondingly had less effective, longer intervals with poorer ratios.  This was because when the mechanism behind a transient was well understood, such as an overflow in the digital control channels of the interferometer, it was easier to identify the specific times at which these transients occurred.  In some cases, however, well understood vetoes may include little enough time that statistics are difficult to perform due to the small number involved, and these can still be categorized as well understood, providing sufficient evidence for coupling is present.  Conversely, when only the general cause was known, as in the case of transients related to passing trains at the Livingston site, long intervals of time when these conditions were present needed to be vetoed in order to capture the related transients, despite the short duration of each particular transient.  In this latter case, a statistical argument based on the veto metrics was required to prove the utility of a set of veto intervals.  

The idea of this categorization scheme was to allow the followup of candidate triggers after applying sequentially each category of vetoes with the consequently lowered background false alarm rate, in order to search for detections~\cite{Collaboration:2009tt,Abbott:2009qj}.  In the CBC searches in LIGO's S5 science run, we decided on four categories in descending order of understanding of the problems involved:

\subsection{Category 1}
The first category includes vetoed times when the detectors were not taking data in the design
configuration.  A fundamental list of science mode times is compiled for each interferometer, and only the data in these times is analyzed.  These times are logged automatically by the detectors with high reliability, though on rare occasions DQ flags marking non-science mode data mistakenly marked as science quality need to be generated after the fact.  These are the same for all searches, and do not need to be padded with extra windows, as the data are not analyzed.

\subsection{Category 2}
The second category contains well understood vetoes with well tuned time intervals, low deadtimes and a firm model for the coupling into the gravitational wave channel.  For many transients, this results in a high efficiency, particularly at high $\rho$, though this is not necessarily the determining factor in categorization.  A well understood noise coupling into the gravitational wave channel may consistently produce triggers of moderate amplitude, or at a lower rate than more common transients.  These are still considered to be of category 2 if several conditions are met.  The ratios $R_{ED}$ and $R_U$ should be statistically significant, of order 10 or higher, for all clusters above some $\rho$ threshold characteristic of the transients.  
\begin{figure}[ht]
\begin{center}
\includegraphics[width=0.5\textwidth]{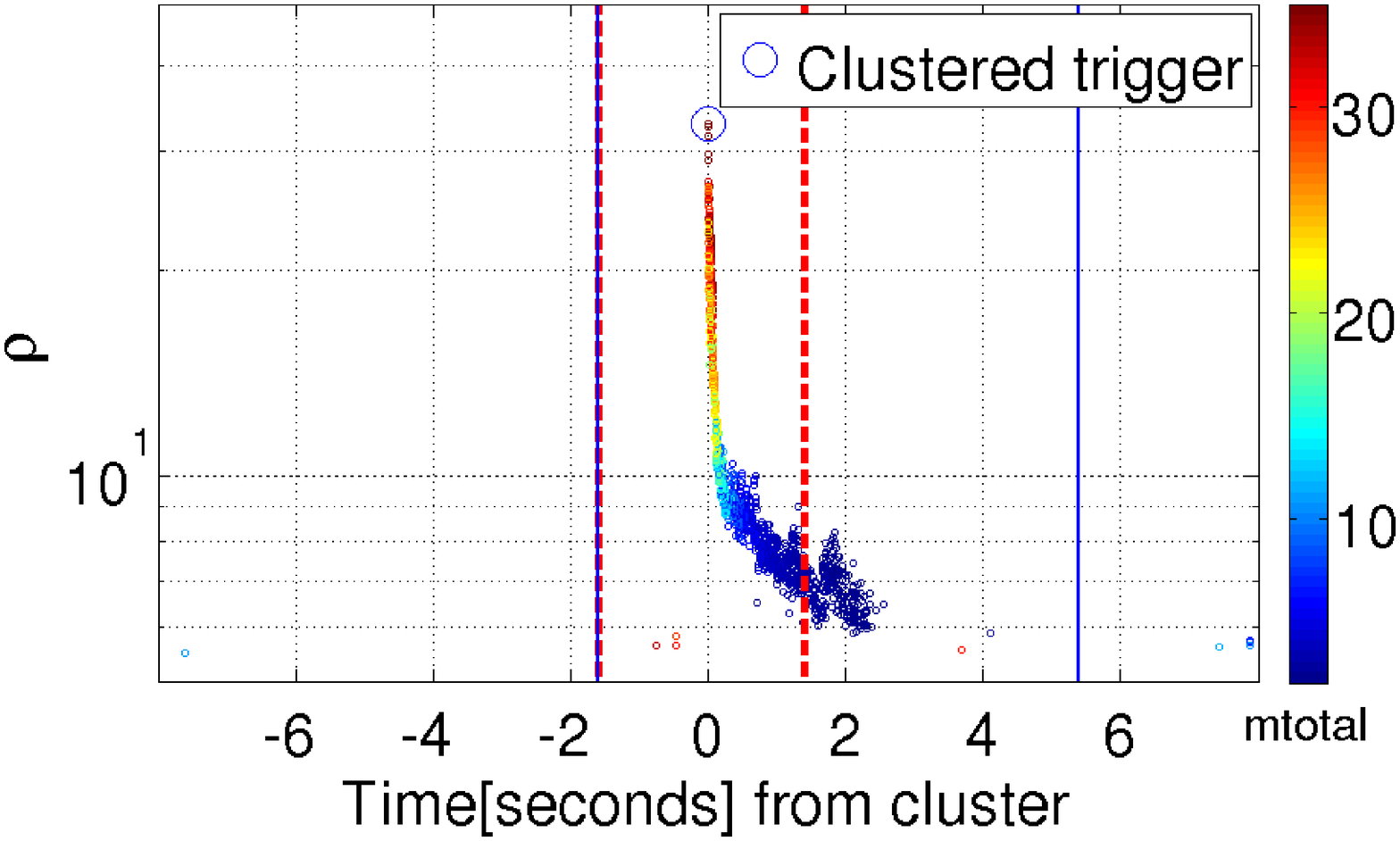}\includegraphics[width=0.5\textwidth]{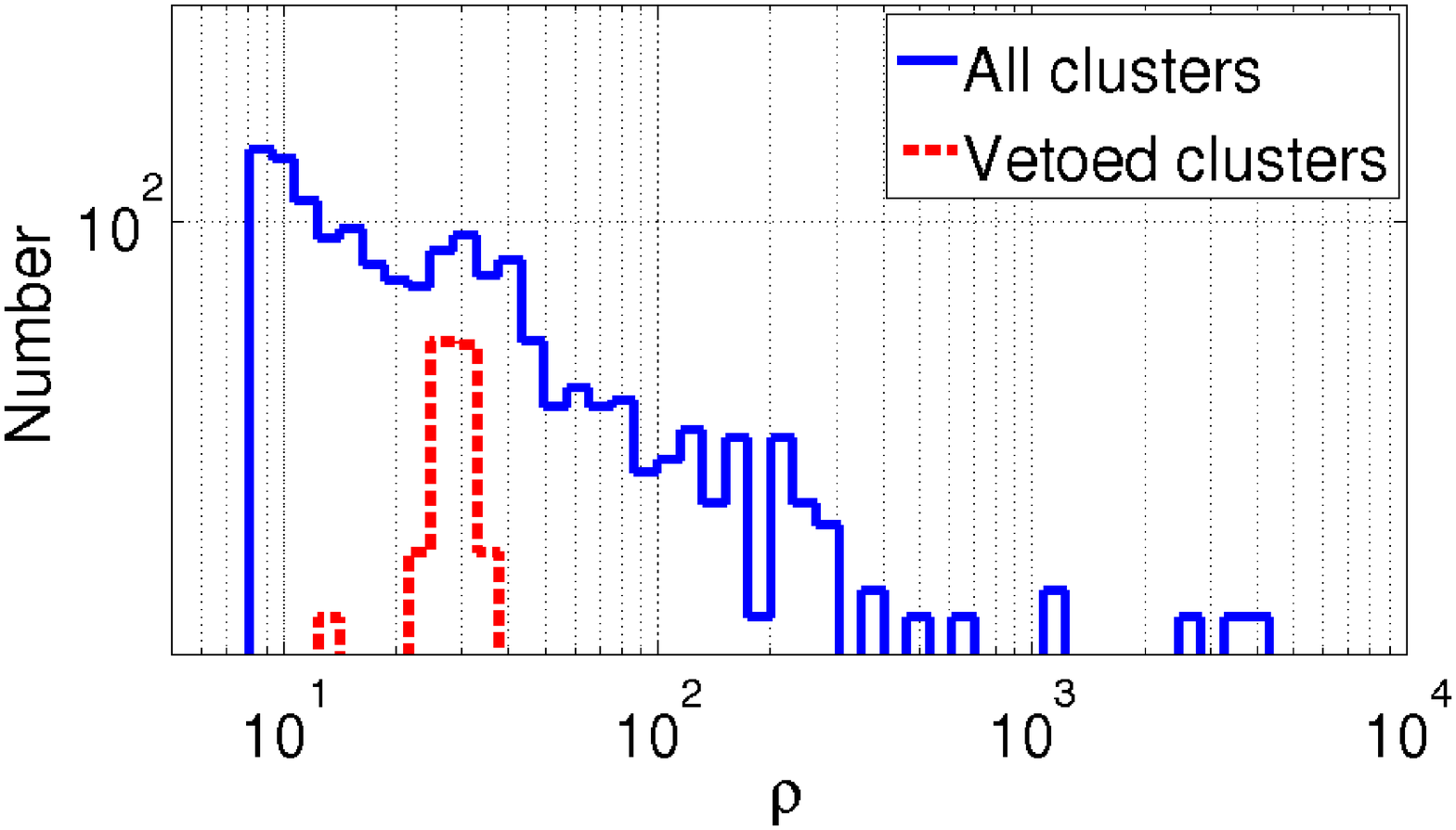}
\end{center}
\caption{ A category 2 data quality veto for a month containing glitches in the TCS lasers of H2.  At right, a log-log histogram of single interferometer clusters for the CBC search with total mass between 2 and 35 solar masses, with all clusters in blue and vetoed clusters in red.  At left, the effect of the transient on the production of unclustered triggers (dots), the 10 second clusters of these raw triggers (circle), the original Data Quality flag (dashed lines), and the expanded data quality veto after the window paddings are applied (solid lines).
}
\label{fig:H2TCStrig}
\end{figure}

One example of a category 2 veto is the overflow veto mentioned in Section 4.3.  As is clear from
the right hand plot in Figure \ref{fig:H2eff}, these veto intervals include most of the loudest
clusters.  Not all category 2 vetoes need to have this level of efficiency, or any efficiency at
all at the most extreme $\rho$. For instance, we also used vetoes based on data quality flags for glitches in the lasers for the thermal compensation system (TCS).  TCS heats the mirrors in order to offset changes in curvature due to heating by the main laser.  These flags, with 4 seconds added to the end of the original 2 second intervals, had a $R_{ED}$ ratio of nearly 500 for triggers above $\rho$ of 20, but zero efficiency above $\rho$ of 40.  The $R_U$ for this veto was over 100.  As is clear in Figure \ref{fig:H2TCStrig}, there is a population of clusters with $\rho$ from 20 to 40 that correspond to the transients from TCS glitches in this particular search. 

\subsection{Category 3}
The third category contains vetoes which were significantly correlated with transients, but with less understanding of the exact coupling mechanism, and thus often poorer performance in the metrics of deadtime and used percentage than category 2 vetoes.  There are many sources of transient noises whose coupling is only partly understood.  Site-wide events of significant duration, such as heavy winds or elevated seismic motion, intermittently lead to loud transient noises.  The auxiliary channel vetoes discussed in Section 4.3, because of their statistical nature, fall also into this category.  Category 3 vetoes also include the minutes immediately preceding the loss of lock of the interferometer, when the triggers were likely due to the same instabilities that contributed to the lock loss.  These vetoes, based more on the probability of transients than a direct measurement, tend to have lower used percentages, higher deadtimes, and therefore smaller ratios between the efficiency and deadtime.  

The train data quality flag veto mentioned in Section 4.2 was in this category, for while the trains themselves were well understood, the nonlinear coupling to create sporadic high frequency glitches was not.  This caused large windows defined by the presence of heightened ground motion alone to be created, rather than targeted vetoes of the individual noise transients.  

\begin{figure}[ht]
\begin{center}
\includegraphics[width=0.5\textwidth]{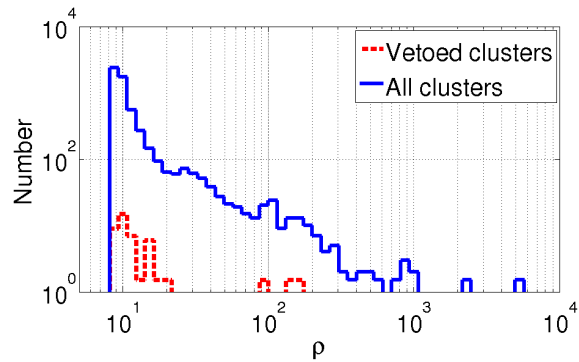}
\end{center}
\caption{ A category 3 data quality veto for a month of high winds at Hanford.  A log-log histogram of single interferometer clusters, total in solid lines and vetoed in dashed lines.  
}
\label{fig:H2Wind trig}
\end{figure}

Another example was elevated winds above 30 Mpc at the Hanford site.  This data quality veto had a $R_{ED}$ of 17, and a $R_U$ of 31.  While this is significant, it is less than the typical value of category 2 vetoes.  

By definition, the auxiliary channel vetoes have a large used percentage, always greater than 50\%.  However, they are classified as category 3 because the coupling is not well understood.  An example of such a veto made using the above technique based on an interferometer control channel is the veto made from the H1 feedback loop that keeps the recycling cavity resonant.  This veto had an $R_{ED}$ of 20.  Another veto, this time based on an environmental channel, used the magnetometers located at the end of the L1 Y-arm, and had an $R_{ED}$ of 10.

\subsection{Category 4}
The fourth category contains vetoes with low statistical significance, often with high deadtimes.  The used percentages are often near 100\%, but this is a representative of the long intervals of science mode time flagged, and thus the high probability that at least one cluster will be within the time defined (as mentioned earlier, the average clustered trigger rate was of order one per 10 minutes).  Seismic flags with lower thresholds, aircraft passing within miles of the detectors, and problems recorded in the electronic logbooks at the detectors all fall within this category.  These long intervals are not used as vetoes for searches, but rather are identified for the purposes of providing input to the follow up of gravitational wave candidates, when all possible factors that prompted the creation of these flags at the time of the detection candidate are scrutinized.  

\subsection{Examining candidate events after vetoing times}

\begin{figure}[ht]
\begin{center}
\includegraphics[width=0.75\textwidth]{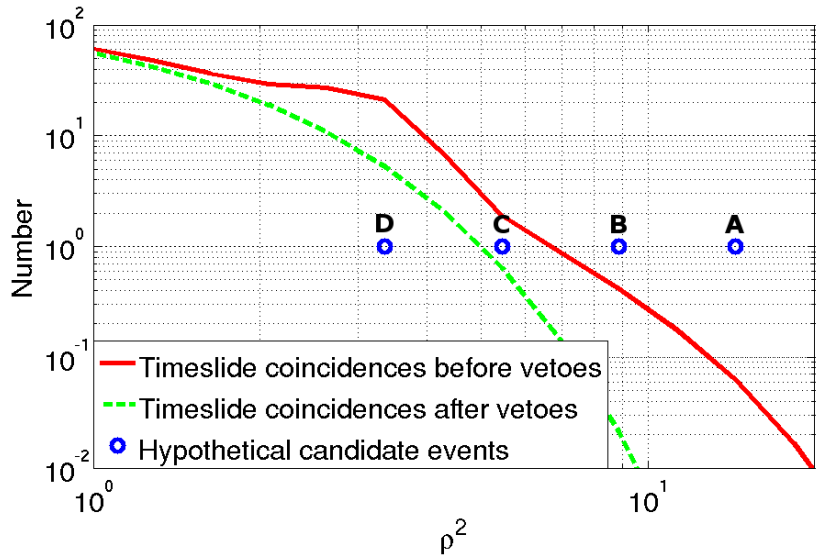}
\end{center}
\caption{This diagram schematically represents the anticipated effect of vetoes on the significance of candidate events.  The solid lines represent the estimated background coincidence rate from timeslides before and after vetoes.  The circles denote the number of foreground coincidences with $\rho^2$ equal to or greater than the x axis value.  For the purpose of the discussion in Section 5.5, the points A, B, C, and D are denote hypothetical detection candidates.
}
\label{fig:coincidentTriggers}
\end{figure}
As mentioned earlier, candidate coincident events that occur during the times of category 2, 3, and 4 vetoes are not automatically discarded.  The total deadtime of category 2 vetoes for the low mass CBC search, for example, was of order 1\%. Category 3 had a deadtime of order 5\%, and category 4 many times that.  As the search reported no detections, it therefore included a calculation of the upper limit on the number of compact binary coalescences.  The first three categories, including both data quality flag vetoes and auxiliary channel used percentage vetoes, were applied in this example search before calculating the upper limit, as they reduced the false alarm rate from these transients.  Because the total veto deadtime of the applied categories was between 5\% and 10\% per interferometer, the probability that a true gravitational wave could be in a vetoed interval is significant.  

The decisions on which vetoes to use are made prior to examination of candidates.  While the veto choices were tuned on single interferometer triggers, the end product of the CBC searches are detection candidates found in coincidence in the data from multiple interferometers.  In the rest of this section, we will discuss the effect that the veto categories have on coincident CBC searches.

While we are unwilling to precipitously remove all candidates in vetoed times from consideration, it is imperative to reduce the rate of accidental coincidences from the noise transients that have been identified. This can be done by examining all significant candidates with respect to the background present after each category is consecutively applied.  If a candidate is not vetoed by successive veto categories, it becomes more significant, as more of the background false alarm candidates against which it is compared are removed.  If a candidate is in vetoed time for a given category, it is not completely ruled out as a detection, but further investigation would be required to show such a candidate is not an artifact due to the disturbance that triggered the veto.  Candidates that are vetoed by lower numbered categories are more suspect, given the firmer understanding of the vetoes that populate the first two categories.  A true gravitational wave is not impossible, as a sufficiently nearby (within the Milky Way) binary system coalescence could theoretically overflow the feedback control systems.  It would be apparent, however, in follow up investigations as the spectrogram for the data would show a large amplitude chirp signature leading up to the overflow, coherently between multiple detectors.

The diagram in Figure \ref{fig:coincidentTriggers} illustrates the effect of vetoes that we anticipate on the significance of detection candidates.  Superimposed are four hypothetical candidates with the labels A, B, C, and D.  For illustration only, let us assume that there is a gravitational wave candidate at one of these points.

If the candidate is at point A, it is visible above background and significant before any vetoes are applied.  If A is not vetoed after subsequent veto categories are applied, it will be the loudest candidate and significantly above the rest of the distribution, and thus a strong gravitational wave candidate.  If A is vetoed, it would be plausible to believe it could still be significant with strong evidence that it did not originate from the same problems used to define the veto, as in the hypothetical example of the galactic coalescence mentioned above.  Even if A is recovered, it would be compared to the background estimation before vetoes are applied, and thus have a lower significance than had it survived the vetoes originally.   

If the candidate is at point B, it is visible above background, though not as significant as if it were at point A.  If it is not vetoed, then B is a good candidate, and having cleared away the understood accidental triggers from transient noises, it can be followed up in depth.  The reason for defining vetoes is precisely to uncover these candidates, which would be buried in the background otherwise.  If it is vetoed, it is again necessary to confirm that the data artifacts prompting the veto intervals are not responsible for the candidate.

If the candidate is at point C, it is likely among the triggers with the largest $\rho^2$ before the vetoes are applied.  Surviving the vetoes improves its ranking, reducing the background of triggers with equal or lower $\rho^2$.  If it is vetoed, it is a problematic candidate given the population of spurious triggers that it sits in.  If it survives the vetoes, it is still only a marginal candidate.  Follow up analysis of the highest $\rho^2$ triggers will likely uncover reasons to distrust the surrounding loud candidates, but that is not enough to make C into a strong candidate.  Additional veto definitions and revisions would make a candidate at C somewhat more significant, providing it is not vetoed, plausibly making it significant.

A candidate at point D is within the accidental population of triggers after vetoes are applied, with tens of triggers surrounding it with similar $\rho^2$.  Such candidates are not detectable without additional reduction of the background.  Additional veto definitions and revisions might make a candidate at D marginally more significant, but it is not at all likely to become detectable through veto efforts.

\subsection{Results of veto efforts in S5}
In S5, hundreds of data quality flags and auxiliary channels were evaluated as mentioned above in Sections 4 and 5.  The resulting veto metrics and categorization for each of the vetoes used in the searches were archived in a technical document~\cite{S5VetoTechNote}.

\section{Proposed Veto Techniques}
The techniques mentioned in Section 4 were by and large refined months after the data were recorded.  Additionally, the decisions on categorization, veto window padding, and utility of auxiliary channels for used percentage vetoes were determined largely by individual human examination of the behavior of each data quality flag and auxiliary channel.  While this was necessary for development and early implementation, the low latency and rigor associated with automating as much of this decision making process as possible is desirable.  Below are discussions of our current and near future efforts to realize the goal of automated evaluation, categorization, and extension of data quality and auxiliary channel vetoes.

\subsection{Automated categorization using a $\chi^2$ test}

One promising method that has been developed, and that has the potential to help automate
recommendations for veto categories, is a figure of merit based on a $\chi^2$ test. For
each DQ flag this $\chi^2$ statistic is given by
$$\chi^2(\rho)=\sum_{k=1}^{R}\frac{(n_k (\rho)-T_k\langle n_t(\rho)\rangle)^2}{T_k\langle
n_t(\rho)\rangle}\,,$$
where $\langle n_t(\rho)\rangle$ is the average number of triggers per unit time
in the science run above a certain threshold $\rho$, $T_k$ is the duration of the flagged
window $k$, and $n_k$ is the actual number of triggers above the threshold $\rho$ in the
same window.

The null hypothesis is that the triggers are Poisson distributed, i.e.\ there is no
correlation between the presence of triggers and the DQ flags.  In our analysis we compute the
figure of merit and test the null hypothesis at a confidence level of 95\%.  The higher the figure
of merit, the higher the correlation between triggers and DQs and thus the lower the category.

As an example of the $\chi^2$ categorization scheme, the $\chi^2$ value for the H2 overflow mentioned previously is $\approx 100$ times higher than the typical ranges of category 3 vetoes and $\approx 1000$ times higher than the ranges typical of category 4 vetoes.  For vetoes that have $\chi^2$ values that are near or on the boundaries between the categories we turn to figure of merits mentioned previously such as deadtime, used percentage, and efficiency to determine which category a particular veto belongs in.  For example, the veto for glitches in the TCS lasers of H2 has a chi-squared value that falls near the lower range for category 2 and the higher range for category 3. However, the high used percentage and the low deadtime distinguish this veto from category 3 vetoes with similar $\chi^2$ values.

The $\chi^2$ method is a step towards automated categorization.  An automated monitor that
incorporates previously mentioned figures of merit and $\chi^2$ values to organize vetoes into
categories is currently being developed. 

\subsection{Automated veto window padding determination}

The determination of window paddings (see Section 4.2) is another step of the veto selection
pipeline that currently requires human input. A method to automate this step would be to look for
quiet time intervals of pre-determined duration around the clusters. If the unclustered triggers
remain below the minimum $\rho$ threshold for the chosen duration before and after the glitch's
loudest trigger, the earliest and latest triggers above the threshold would determine the left and
right padding of the DQ window, respectively. Assuming the duration of the required padding for
the DQ flag to be normal distributed, a final recommendation for the padding of that flag would be
obtained by taking the average of the values for each window.

\subsection{Multiple auxiliary channel veto algorithm}

Another approach that has been explored is that of defining vetoes when multiple auxiliary channels glitch coincidently, specifically by examining the output of the ``QScan'' time-frequency algorithm~\cite{Chatterji:2004qg} over multiple auxiliary channels, at the time of detection candidates.  When a number of auxiliary channels glitch simultaneously and at the same time as the gravity wave channel, there is a strong possibility that the glitching has a non-astrophysical cause, particularly when the channels are physically related.  For example, a number of the length sensing and control channels may glitch together in different parts of the interferometer, such as the beam splitter and reflected and dark ports.  Sometimes the glitches in a set of length sensing and control channels will be associated with glitches in the alignment sensing and control channels.  Many of these measure pitch and yaw of mirrors, including the test masses, and when mirror alignment and length disturbances occur simultaneously, it is unlikely that a transient also in the gravity wave channel will be astrophysical.  Nevertheless, safety studies have been successfully conducted, using all of the hardware injections from the first year of S5, to verify that these combinations of glitches could not be caused by the arrival of gravitational waves.  
	
Tests of the efficacy of proposed vetoes were carried out on data from the S5 search with total mass between 25 and 100 solar masses.  On outlier coincident triggers with $\rho \geq 200$ remaining after the application of existing veto categories 1 through 4, from 94 to 100\% of the triggers would be vetoed for each single interferometer, and 100\% of the coincident triggers would be vetoed in one or more interferometers.  Since these vetoes are run on small time intervals around the times of the detection candidate triggers, the dead time is not calculated: a candidate survives or is vetoed.  

Future work planned includes running this algorithm over the times of candidates using a larger set of the available channels, rather than only channels corresponding the length and alignment sensing and control, as restricted in the above studies to reduce computation time.  It is expected that additional sets of channels will be useful, particularly the environmental channels.

\section{Summary}

In this paper, we showed how we developed techniques for vetoing non-astrophysical transient noises safely and effectively, in order to reduce the effect of noise transients on astrophysical searches for low mass CBCs in the first year of the initial LIGO data run. We based our vetoes on data quality flags created by detector characterization work, as well as KW triggers from auxiliary channels with high used percentages.  Though we approached each flag and each channel individually, and though different flags and different channels reflected a variety of specific causes, we found that the effects on the gravitational wave channel fell into a few common groupings.  Flags and channels that responded to similar phenomena generally required similar windows, had similar deadtimes, were effective for similar populations of triggers, and therefore were placed in the same categories.  The LSC used these categories for sequentially studying the significance of gravitational wave candidates rising above background in the CBC searches. 

Going forward, we intend to use the experience gained to finish ongoing automation work both to select veto window paddings and to provide recommendations for veto categorization.  There will be data quality flag and auxiliary channel based vetoes developed for use in LIGO's S6 science run. The goal will be to analyze the auxiliary channel KW triggers in near real time, and to have vetoes defined on a week-by-week basis for both types of vetoes.  It is probable that there will be marginal cases for categorization that require further human review, but automation will allow us to focus our time on these cases.  

\ack
This work is partially supported by the National Science Foundation grants PHY-0457622 (NZ, TR), PHY-0553422 (NC, TI, MC, JC), PHY-0555406 (KR), PHY-0600259 (JRS), PHY-0605496 (GG, JS), PHY-0653550 (LC), PHY-0757937 (MC, BR), PHY-0757957 (PS), PHY-0847611(DAB), PHY-0854790(NC, TI, MC, JC), and PHY-0905184 (GG, JS).

\section*{References}

\providecommand{\newblock}{}

\end{document}